\documentclass[12pt,titlepage]{article}
\usepackage{amsmath}
\usepackage{amssymb}
\usepackage{graphicx}
\usepackage{caption2}
\usepackage{amsfonts}
\usepackage{cite}
\usepackage{lineno}

\newcommand{\be}{\begin{equation}}
\newcommand{\ee}{\end{equation}}
\newcommand{\bea}{\begin{eqnarray}}
\newcommand{\eea}{\end{eqnarray}}
\newcommand{\bef}{\begin{figure}}
\newcommand{\ef}{\end{figure}}
\newcommand{\bt}{\begin{tabular}}
\newcommand{\et}{\end{tabular}}
\newcommand{\bno}{\begin{enumerate}}
\newcommand{\eno}{\end{enumerate}}

\def\3{\ss}

\catcode`\"=\active
\def"{\accent'177}
\begin{document}

\begin{center}

{\bf On the inherent self-excited  macroscopic  randomness  \\ of chaotic three-body systems}

Shijun Liao $^{a,b}$  and   Xiaoming  Li $^b$ 

$^a$ Ministry-Of-Education  Key Laboratory in Scientific Computing\\
$^b$ State Key Laboratory of Ocean Engineering \\  
School of Naval Architecture, Ocean and Civil Engineering \\
Shanghai Jiaotong University, Shanghai 200240, China\\
E-mail:  sjliao@sjtu.edu.cn

\end{center}

{\bf Abstract}
 What is the origin of macroscopic randomness (uncertainty)?  This is one of the most fundamental open questions for human being.  In this paper, 10000 samples of reliable (convergent), multiple-scale (from $10^{-60}$ to $10^2$)  numerical simulations of a chaotic three-body system indicate that, without any external disturbance, the  microscopic inherent uncertainty (in the level of $10^{-60}$)  due to physical  fluctuation of initial positions of the three-body system  enlarges exponentially into macroscopic randomness (at the level $O(1)) $ until $t=T^*$,  the so-called physical limit time of prediction,  but propagates algebraically  thereafter  when accurate prediction of orbit is impossible.   Note that these 10000 samples use micro-level, inherent physical fluctuations of initial position, which have nothing to do with human being.   Especially, the differences  of these 10000  fluctuations are mathematically so small (in the level of $10^{-60}$ ) that they are physically the same since a distance shorter than a Planck length does  not make physical senses according to the spring theory.   It indicates that the macroscopic randomness  of the chaotic three-body system is self-excited, say, without any external force or disturbances, from the inherent micro-level uncertainty.  This provides us the new concept ``self-excited macroscopic randomness (uncertainty)''.   It is found that  the macroscopic randomness is even dependent upon microscopic uncertainty, from statistical viewpoint.    In addition, it is found that,  without any external disturbance,   the chaotic three-body system might randomly disrupt  with the symmetry-breaking  at $t=1000$ in about 25\%  probability,   which provides us the new concepts ``self-excited random disruption'', ``self-excited random escape'' and ``self-excited symmetry breaking'' of the chaotic three-body system.  It  suggests that a chaotic three-body system might randomly evolve by itself, without any external forces or disturbance.
Thus, the world is essentially uncertain,  since such kind of self-excited  macroscopic randomness (uncertainty) is inherent and unavailable.     This work also implies that  an universe could  {\em randomly} evolve by {\em itself} into complicated structures,  {\em without} any external forces.  To emphasize this point, the so-called ``molecule-effect'' (or ``non-butterfly effect'')  of  chaos  is suggested in this paper.  
  All of these reliable computations could deepen our understandings of chaos from physical viewpoints,  and reveal a kind of origins of macroscopic randomness/uncertainty.


\setlength{\parindent}{0.75cm}

\section{\label{sec:introduction}Introduction }

When one looks at the sky in a clear night,  he/she would feel that stars seem to distribute randomly.   Besides,  velocities in turbulent flows are always different even at the same points of observation using the same measure equipments.   Indeed,  random/uncertain macroscopic phenomena happen quite frequently in practice.  However,  what is the origin of macroscopic randomness (uncertainty)?  This is one of the most fundamental questions for us.   Without doubt, the answers to this open question may greatly deepen and enrich our understandings about nature.

It is widely accepted that microscopic phenomena are essentially uncertain, although they can be well described  by deterministic laws in statistic meanings.   Are there any relationships between microscopic uncertainty and macroscopic randomness?   Some believe that there should exist relationships between them, but some categorically deny.  However, neither of them can provide scientific supports based on  validated  experiments and/or reliable numerical simulations.     

It seems very difficult to reveal the relationship between micro-level and macroscopic uncertainty by means of physical experiments,  because artificial uncertainty of physical experiments caused by human being  is  often  much larger than inherent micro-level physical uncertainty.

Fortunately, it is widely believed that  the  characteristics of nature can be well described by physical laws and principles that are expressed by mathematical formulas/equations.  Like physical experiments,  studies on mathematical models also make great contributions for us to understand the nature better.   For example,  Galileo's and Einstein's famous ``ideal experiments''  completely  renewed  our concepts about inertia, gravity, time and space.

However, there also exists the uncertainty of theoretical prediction using mathematical models, too.   For example, the imperfection of initial condition and numerical algorithms caused by human being might be the sources of uncertainty for prediction.    These sources of uncertainty, caused by limited  accuracy of measurement for initial/boundary conditions and numerical errors of algorithms, are artificial.  But, some sources of uncertainty are physical and inherent, which have nothing to do with human being.  These inherent physical uncertainties are unavailable, although they are much smaller than the artificial uncertainties.  It is a  pity  that  they were curtly neglected in the past.

Generally speaking,  it is difficult to accurately simulate propagation of uncertainty, especially for
chaotic dynamic systems far from equilibrium state,  which have the so-called sensitive dependence on initial conditions (SDIC) \cite{Lorenz1963, Lorenz2006, Wang2012, Logg2013}, i.e. a small disturbance in initial condition leads to huge difference of solution (trajectory).  This is mainly because the artificial numerical noises (truncation error and round-off error) are unavoidable at each time-step, which   enlarge exponentially and propagate together with uncertainty of initial conditions.   In general, uncertainty of initial condition caused by imperfect and limited measurement is often larger than numerical noises, and numerical noises are much larger than inherent, micro-level physical uncertainty of initial condition.   This might be the reason why  inherent, micro-level physical uncertainty of initial condition was hardly considered in a macroscopic chaotic dynamic system.

 In 2009, the method of  ``Clean Numerical Simulation''  (CNS) \cite{Liao2009}  was proposed to   decrease numerical noises so  greatly  that  numerical  errors  can  be    much smaller  even  than micro-level inherent physical uncertainty of initial conditions  in a given interval of time, and thus can be neglected.  The CNS \cite{Liao2009, Liao2013, Liao2013-3b, Wang2012, LiaoWang2014, XiaoMingLi-2014, Liao2014-IJBC}  is based on an arbitrary-order Taylor series method (TSM)  \cite{Corliss1982, Barrio2005} and the arbitrary multiple-precision (MP) data   \cite{Oyanarte1990}, together with a check of solution verification.    Currently, assuming that the initial conditions are exact, a reliable convergent chaotic solution of the famous Lorenz equation in a rather long time interval [0,10000] was gained \cite{LiaoWang2014}, for the first time,  by means of  the National Supercomputer TH-1A (at Tianjing, China),  which is No 17 in the list of top 500 ( http://www.top500.org/list/2014/11/ ),  and the CNS with the 3500th-order Taylor series expansion and the 4180-digit multiple precision data.    It indicates that, given an {\em exact} initial condition,  one can obtain reliable (convergent) solution of  a chaotic dynamic system in {\em any} a finite interval of time \cite{Liao2014-IJBC}, {\em without} uncertainty.   These reliable chaotic simulations in such a long interval of time are helpful to stop the intense arguments about chaos \cite{Liao2014-IJBC}.   It suggests that, for chaotic dynamic systems,  uncertainty might come from initial condition only, since numerical noises can be neglected.

Unfortunately,  the uncertainty of initial condition is unavoidable, not only due to imperfection and finite accuracy of measurement but also due to the micro-level inherent {\em physical} uncertainty.   Traditionally, most researchers often add a small disturbance (which is much larger than micro-level physical uncertainty)  to initial conditions, but without considering its source.   Note that uncertainty is a characteristic of nature, and thus should have nothing to do with the existence of human being:   even if  human being could perfectly measure initial condition in arbitrary accuracy, there still exists the micro-level inherent physical uncertainty of initial conditions.

Without loss of generality,  let  us  consider the famous three-body problem governed by Newtonian gravitational law with the dimensionless equations
 \begin{equation}
    \ddot{x}_{k,i}=\sum_{j=1,j\not=i}^3\rho_j\frac{(x_{k,j}-x_{k,i})}{R_{i,j}^3}, k=1,2,3,
\end{equation}
where ${\bf r}_i = (x_{1,i}, x_{2,i},x_{3,i})$ denotes the dimensionless position of the $i$th body,
$ \rho_i={m_i}/{m_1}$ $(i=1,2,3)$ the ratio of mass,  and
\begin{equation}
    R_{i,j}=\Bigg [\sum_{k=1}^3\left(x_{k,j}-x_{k,i}\right)^2\Bigg]^{1/2}.
\end{equation}
We consider here the case $\rho_1=\rho_2=\rho_3=1$.   As long as velocities of each body are much less than the light speed,    this model is rather accurate in physics, since Einstein's general relativity is unnecessary.    Besides,  by  means  of  the  CNS,  the uncertainty due to numerical noises can be negligible.   In this way, the uncertainty due to  physical model and numerical algorithm is negligible.

However, even if we {\em assume} that  we could measure the initial positions ${\bf r}_i(0)$ and velocities $\dot{\bf r}_i(0)$  in infinite accuracy (although this is impossible in practice according to the Heisenberg's uncertainty principle),  the initial positions of each body are still {\em inherently} uncertain in physics.
First of all,  according to wave-particle duality of de Broglie \cite{Broglie1923},  a body has non-zero amplitude of the de BroglieÕs wave so that  position of a body is always uncertain: it could be almost {\em anywhere} along de BroglieÕs wave packet \cite{Feynman1990}. 
Besides,   the so-called Planck length
\[  l_p = \sqrt{\frac{\hbar G}{c^3}} \approx 1.616252(81) \times 10^{-35} \;\; \mbox{(m)}  \]
is the length scale at which quantum mechanics \cite{Mehra1982}, gravity and relativity all interact very strongly, where $c$ is the speed of light in a vacuum, $G$ is the gravitational constant, $\hbar$ is the reduced Planck's constant, respectively.  According to the string theory \cite{Davies1988}, the Planck length is the order of magnitude of oscillating strings that form elementary particles, and {\em shorter length (than the Planck length) do not make physical senses}.  Especially, in some forms of quantum gravity, it becomes {\em impossible} to determine the difference between two locations {\em less} than one Planck length apart.  Therefore, the micro-level inherent fluctuation of position of a body shorter than the Planck length is essentially uncertain and/or random.
It should be emphasized once again that such kind of uncertainty of position is {\em inherent} and {\em objective}: it has nothing to do with human being and Heisenberg's uncertainty principle.

Using the diameter $d_M\approx 10^{20}$ (meter) of Milky Way Galaxy as characteristic length,  we have the dimensionless physical uncertainty of initial position $l_p/d_M \approx 1.8 \times 10^{-56}$, where $l_p$ is the Planck length given above.  So, from the physical viewpoint, any dimensionless distances shorter than $1.8\times 10^{-56}$ have {\em no} physical senses.  Therefore, it is physically reasonable to assume that the initial velocities $\dot{\bf r}_i(0)$ of the three-body system are exact but their initial positions ${\bf r}_i(0)$ contain a micro-level fluctuation ${\bf r}'_i(0)$ in Gaussian normal  distribution with zero mean and standard deviation $\sigma_0=10^{-60}$, i.e.
$  {\bf r}_i(0) = \overline{\bf r}_i(0) + {\bf r}'_i(0)$,
where $\overline{\bf r}_i(0)=\left<{\bf r}_i(0)\right> $,  $\left<{\bf r}'_i(0)\right> =0$ and $\sqrt{\left<{\bf r}'^2_i(0)\right>} =\sigma_0$.    It should be emphasized that, although all of these initial positions ${\bf r}_i(0)$ are {\em mathematically different}, they are the {\em same} from {\em physical} viewpoint, since a spatial distance smaller than one Planck length  does {\em not} make physical senses.

Without loss of generality, let us   consider the case
\begin{equation}
   \overline{\bf r}_1 = (0,0,-1), \overline{\bf r}_2=(0,0,0), \overline{\bf r}_3=-(\overline{\bf r}_1+\overline{\bf r}_2),
\end{equation}
with the exact initial velocities
\begin{equation}
    \dot{\bf r}_1=(0,-1,0), \dot{\bf r}_2=(1,1,0), \dot{\bf r}_3=-(\dot{\bf r}_1+\dot{\bf r}_2).
\end{equation}
When there is no fluctuation, i.e.  ${\bf r}'_i(0)=0$,  the three-body system is chaotic with a positive  Lyapunov exponent  $\lambda=0.1681$ and a symmetry of motion (i.e. the Body-2 moves along a straight line, and positions of Body-1 and Body-3 are symmetric about this line), but {\em without} disruption (i.e. no body escapes),  as pointed out by \cite{Sprott2010}.    In this case, it is a chaotic system near its equilibrium point with symmetry.       

Many researchers investigated three-body problem.  For example, \cite{monaghan1976a, monaghan1976b} and \cite{nash1978} proposed a statistical theory to study the disruption of three-body systems.    \cite{mikkola2007correlation}  and  \cite{urminsky2009relationship} researched the relation between instability and Lyapunov times for three-body problem.   However,  most researchers simply give a small disturbance of initial condition, but without considering the sources of them: these disturbances are much larger than micro-level physical uncertainty of initial condition mentioned above.   This is because they use low-order numerical algorithms just with double-precision data, which are  certainly not accurate enough to  investigate the propagation of  micro-level  inherent physical uncertainty of initial conditions of chaotic systems.

 \section{Results and discussions}

 How about the propagation of the micro-level inherent physical uncertainty of the initial position ${\bf r}_i(0)$ of this chaotic three-body system?

Ten thousand samples of reliable (convergent), multiple-scale (from $10^{-60}$ to $10^2$) numerical simulations of the chaotic three-body system  are obtained in the time interval [0,1000] by means of the CNS using the multiple-precision  data in  300-digit precision, the time-step $\Delta t =10^{-3}$ and high-enough order $M$ of Taylor series expansion, where $M \geq 30$ in general.   Each chaotic solution is verified by means of a higher-order Taylor series expansion with the same initial condition, and only convergent results in the interval [0,1000] are accepted.  In this way,  the artificial uncertainty due to numerical algorithms is negligible, since  numerical noise is much smaller than physical uncertainty in the whole interval of time under consideration.   
Thus, the micro-level inherent physical fluctuation  ${\bf r}'_i(0)$  of the initial position ${\bf r}_i(0)$ is the only source of the uncertainty.

\begin{figure}[h]
\begin{center}
\includegraphics[scale=0.5]{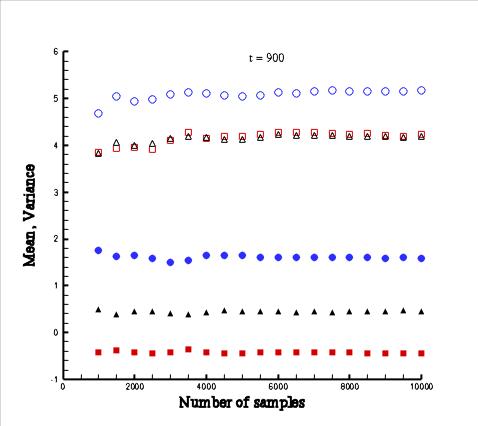}
\caption{{\bf The mean and standard deviations of Body-1 at $t=900$ versus the number of samples.}   Results are  based on the different numbers of  samples of reliable, multiple-scale simulations  given by the CNS with the micro-level fluctuation of initial position ${\bf r}'_i(0)$ in Gaussian distribution ($\sigma_0=10^{-60}$).  Filled  Square: $\overline{x}_{1,1}$;  Filled  Delta: $\overline{x}_{2,1}$;  Filled  Circle: $\overline{x}_{3,1}$;  Square:  $\sigma_{1,1}(t)$;  Delta:  $\sigma_{2,1}(t)$;  Circle:  $\sigma_{3,1}(t)$.}
\label{Fig:mean-sample}
\end{center}
\end{figure}

Let $\overline{x}_{i,j}(t)$ denote the mean of $x_{i,j}(t)$  and $ \sigma_{i,j}(t)$
its unbiased estimate of standard deviation, respectively, based on 10000 samples  of the reliable convergent CNS simulations \cite{Liao2009, Liao2013, Liao2013-3b, LiaoWang2014, XiaoMingLi-2014} using the initial conditions with different micro-level fluctuations ${\bf r}'_i(0)$ of position.   So, we have the initial standard deviation
$ \sigma_{i,j}(0) = \sigma_0=10^{-60}$ for the considered case.   It took one and a half day to calculate these 10000 samples  using  an  account  with 1024 CPUs in the National Supercomputer TH-1A at Tianjian, China, which is No. 17 in the list of top 500 (see http://www.top500.org/list/2014/11/ ).   It is found that both of the means $\overline{x}_{i,j}$ and the standard deviations  $\sigma_{i,j}(t)$  are almost the same as the number  of samples is greater than 8000, as shown in Fig.~\ref{Fig:mean-sample}.  So, it is reasonable for us to use 10000 samples in this paper.

\begin{figure}[h]
\includegraphics[scale=0.4]{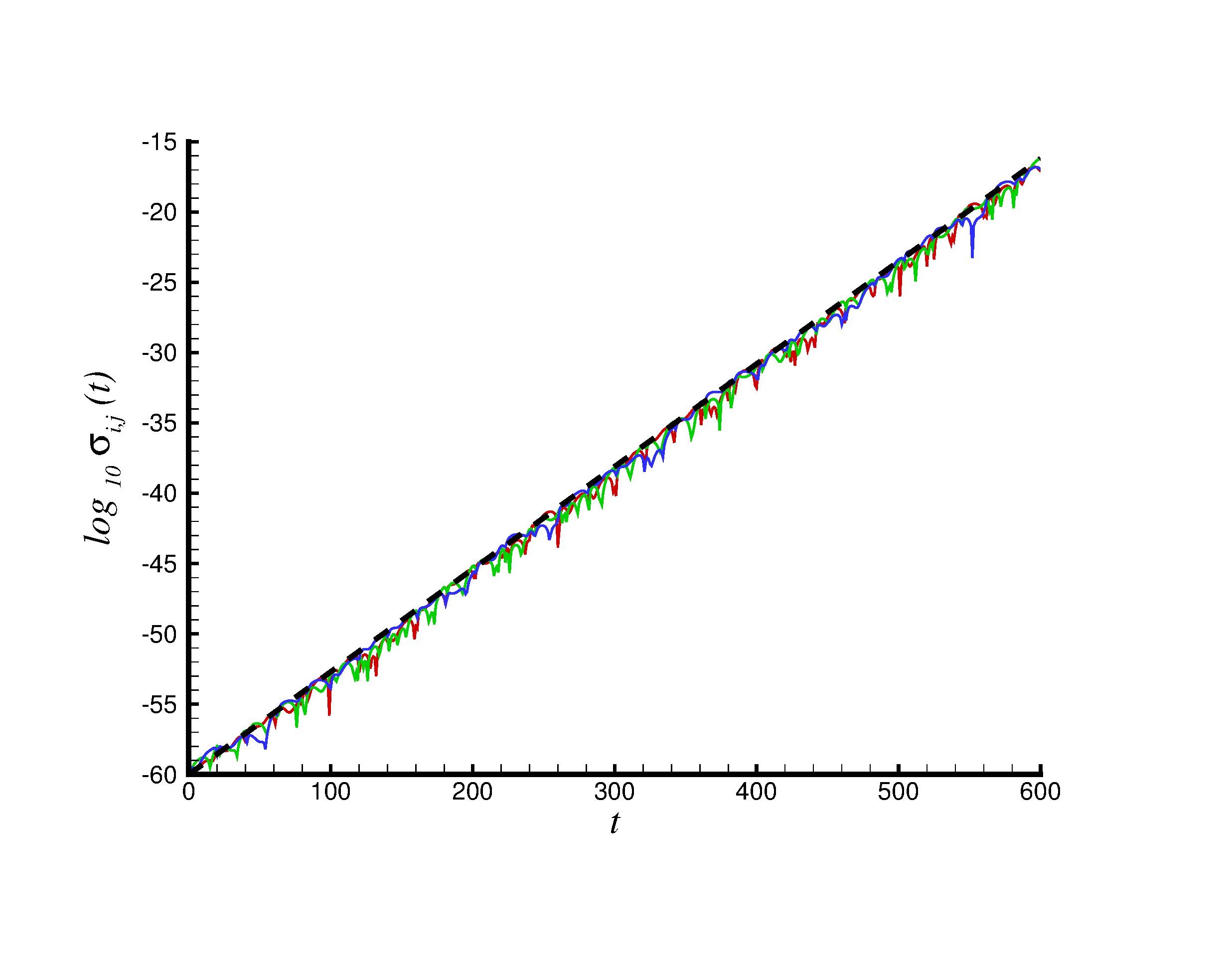}
\includegraphics[scale=0.4]{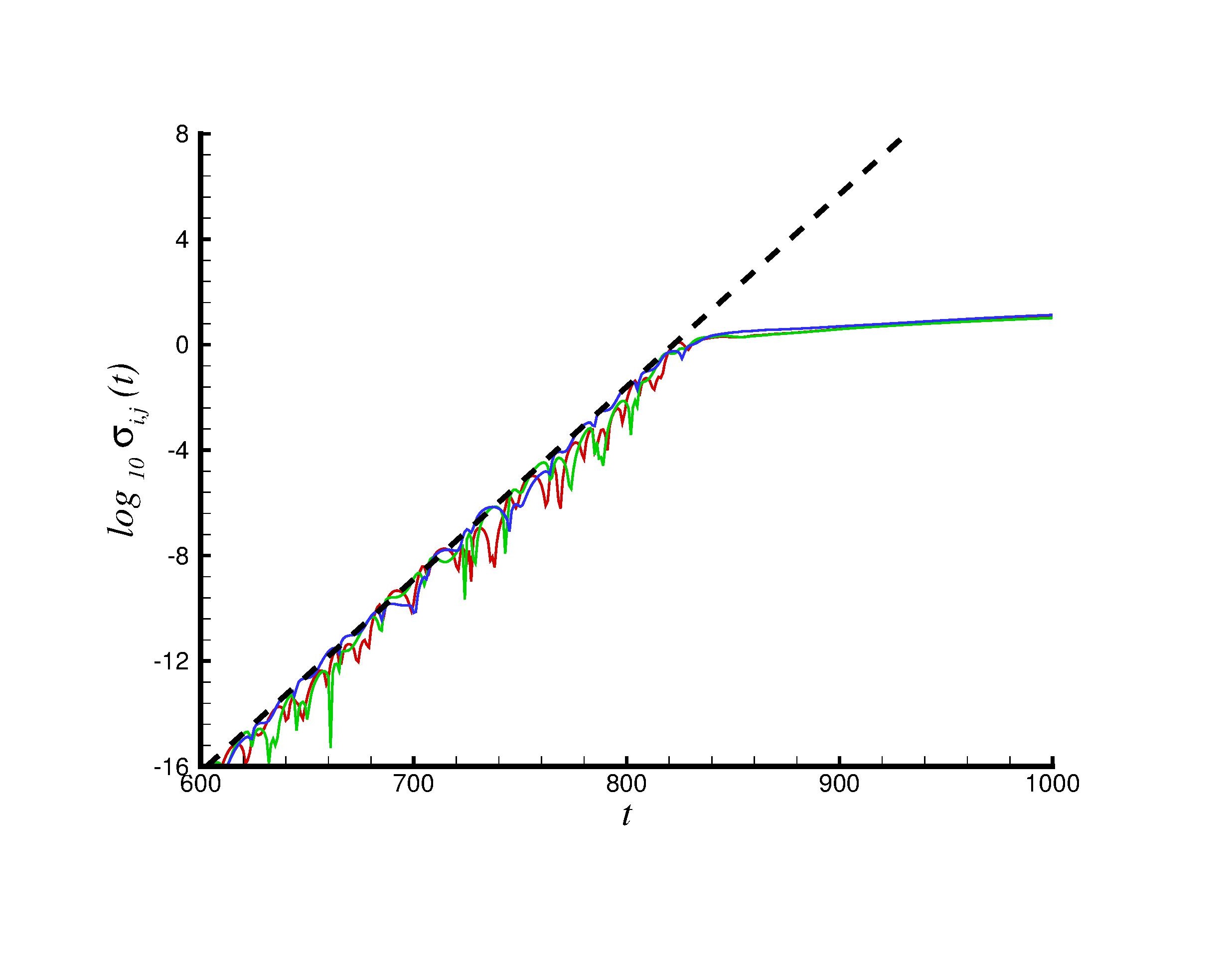}
\caption{{\bf The standard deviations of Body-1.}   Results are  based on the 10000  samples of reliable, multiple-scale simulations  given by the CNS with the micro-level fluctuation of initial position ${\bf r}'_i(0)$ in Gaussian distribution ($\sigma_0=10^{-60}$).    Red line:  $\sigma_{1,1}(t)$;  Green line:  $\sigma_{2,1}(t)$;  Blue line:  $\sigma_{3,1}(t)$;  Dashed line:  $\sigma = \sigma_0 \; exp(\lambda t)$ where $\lambda=0.1681$ is the Lyapunov exponent  given by \cite{Sprott2010}.}
\label{Fig:sigma}
\end{figure}

\begin{figure}[t]
 \begin{center}
\includegraphics[scale=0.3]{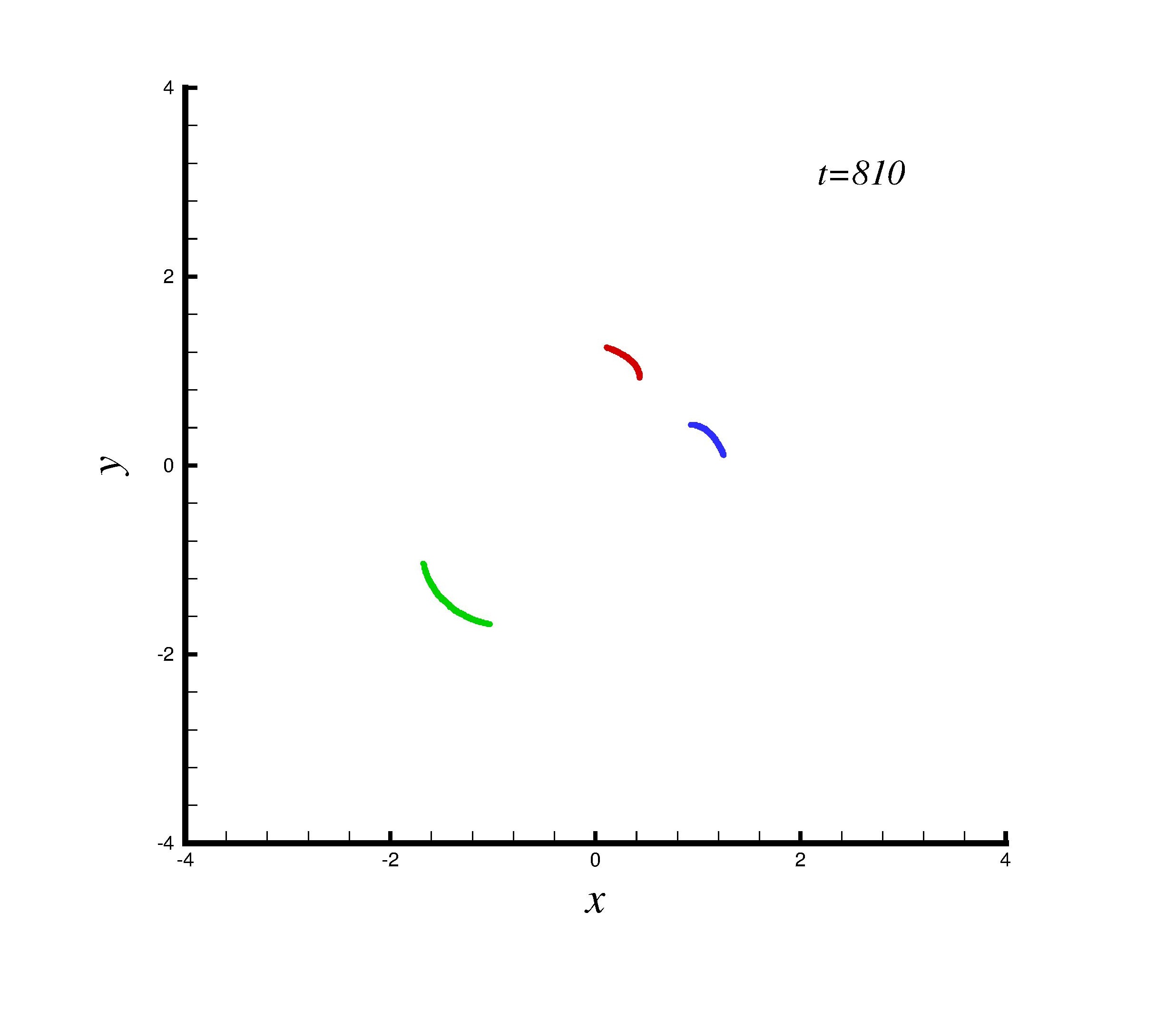}  \includegraphics[scale=0.3]{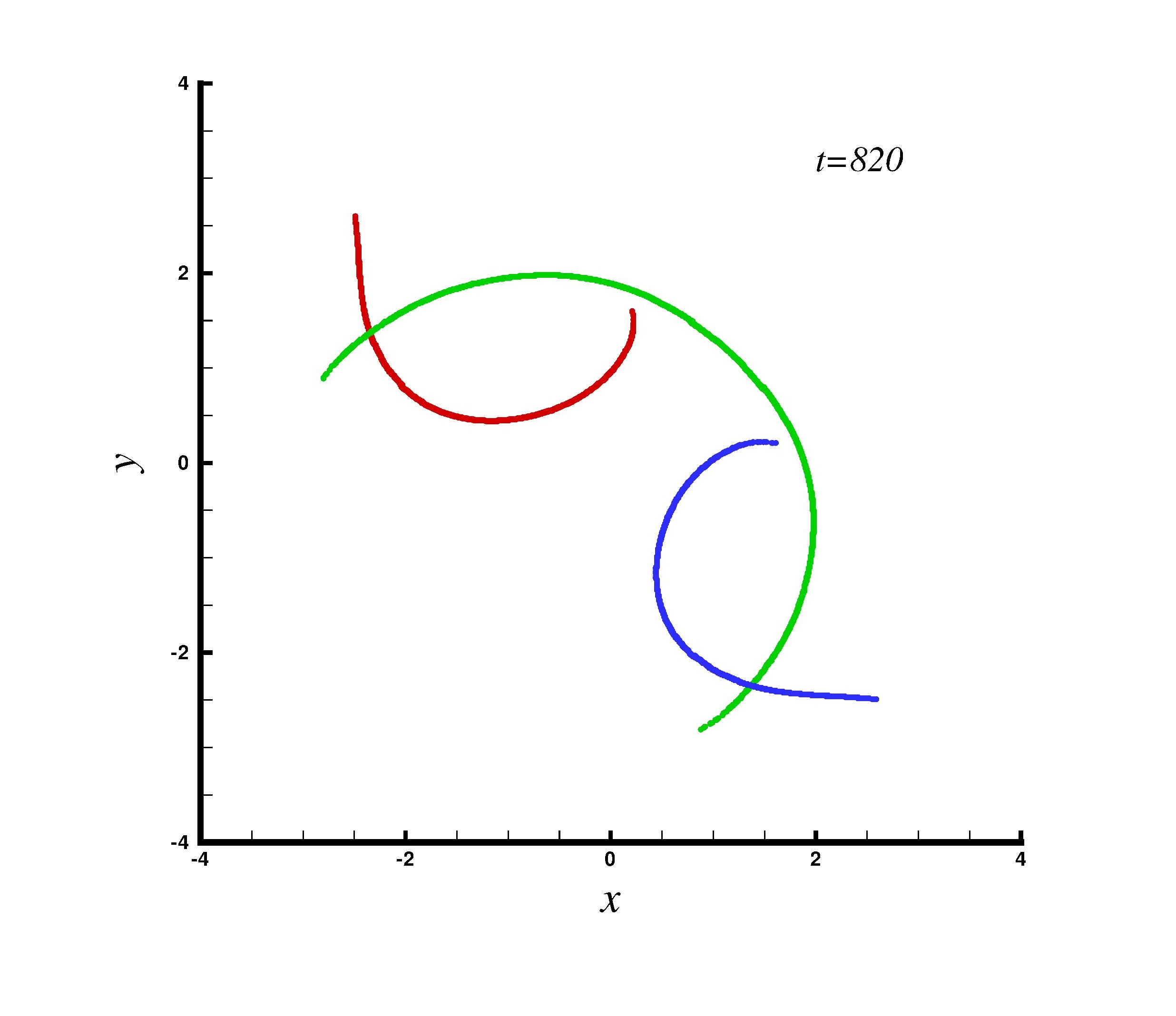} \\
\includegraphics[scale=0.3]{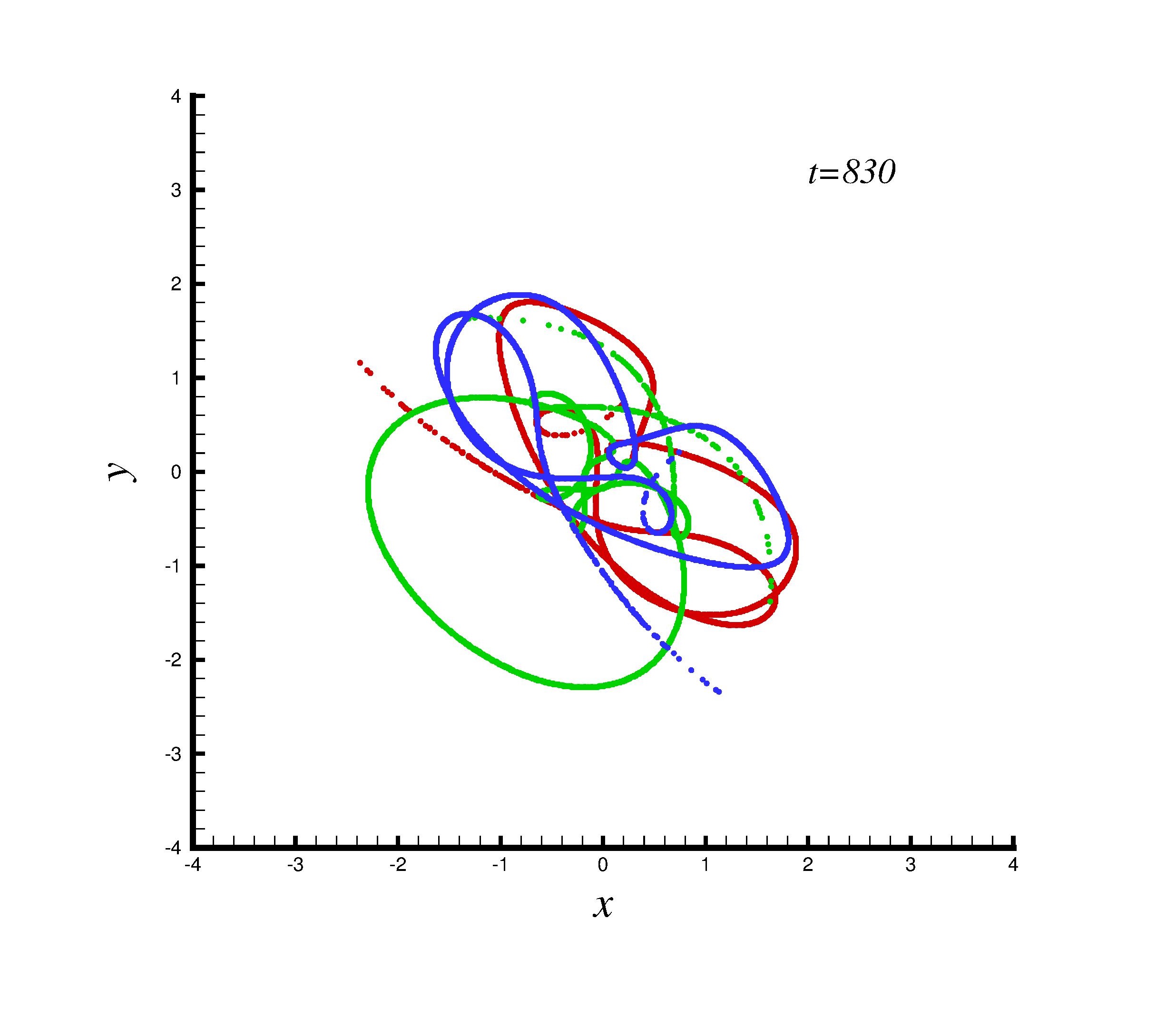}  \includegraphics[scale=0.3]{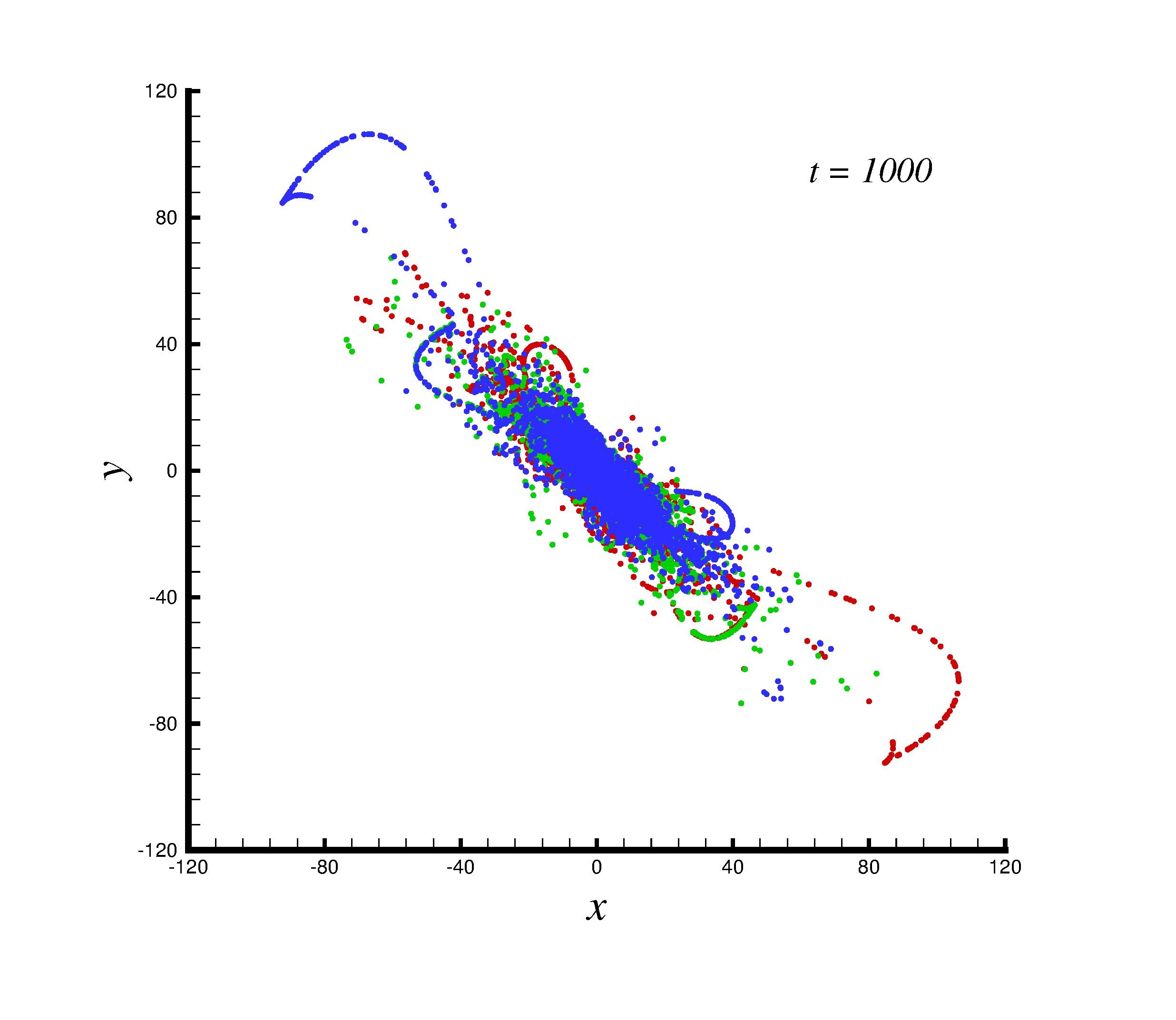}
\caption{{\bf The position distribution of Body-1 (red points), Body-2 (green points) and Body-3 (blue points) in the $(x,y)$ plane at different times when $\sigma_0=10^{-60}$.}  Results are  based on the 10000  samples of reliable, multiple-scale simulations given by the CNS with the micro-level fluctuation of initial position ${\bf r}'_i(0)$ in Gaussian distribution.  {\bf The corresponding two movies are published on the website of the journal}.  }
\label{Fig:transition}
\end{center}
\end{figure}

Obviously, the smaller the standard deviation $\sigma_{i,j}(t)$, the smaller the uncertainty.
According to the statistic analysis based on these 10000  samples of  reliable (convergent)  simulations given by the CNS,  each $\sigma_{i,j}(t)$  increases  exponentially  from $\sigma_{i,j}(0) = 10^{-60}$ when $t=0$ until $\sigma_{i,j}=\sigma^*$ at $t =T^* \approx 810$, as shown in Fig.~\ref{Fig:sigma} for Body-1 as an example,  where   $\sigma^*$ is a standard deviation corresponding to an observable macroscopic difference of position, $T^*$ is the critical time corresponding  to $\sigma^*$, respectively.   It is found that the uncertainty propagates exponentially in essence and can be expressed approximately in the form
 \[  \sigma_{i,j}(t)  \approx  \sigma_{i,j}(0) \; e^{\lambda t} = \sigma_0 \;  e^{\lambda t}, \;\;\; 0\leq t < T^*,  \]
where $\lambda = 0.1681$ is exactly the Lyapunov's exponent for the same chaotic three-body system  \cite{Sprott2010} with the {\em exact} initial conditions
\begin{eqnarray}
\dot{\bf r}_1&=&(0,-1,0), \dot{\bf r}_2=(1,1,0), \dot{\bf r}_3=-(\dot{\bf r}_1+\dot{\bf r}_2),  \label{mean-position:0} \\
{\bf r}_1 &=& (0,0,-1),  {\bf r}_2=(0,0,0), {\bf r}_3=-( {\bf r}_1 + {\bf r}_2),  \label{mean-velocity:0}
\end{eqnarray}
say, without the micro-level fluctuation of position, i.e. ${\bf r}'_i(0) =0$.     Besides, the critical time $T^*$ is approximately determined by
\begin{equation}
\sigma_0 \; e^{\lambda T^*}  = \sigma^*.   
\end{equation}
 For example, one has $T^* =  801$ when $\sigma^* =0.03$,   $T^* =  808.2$ when  $\sigma^* =0.1$,   which  agree  well  with  the observed value of the critical time,  as  shown  in Figs.~\ref{Fig:sigma} and \ref{Fig:transition}.

It is found that,  when $t > T^*$,  the standard deviations $\sigma_{i,j}(t)$ does {\em not}  increase  exponentially  any  more, as shown in Fig.~\ref{Fig:sigma}.    This is a surprise,  since it is traditionally believed that, due to the SDIC, a difference of initial condition of chaotic dynamic systems should be enlarged exponentially.    Note that the similar phenomena  were reported by  Ding and Li \cite{Li2007} for Lorenz equation.     It suggests that $T^*$ is indeed special,  which should have an important physical meaning.

Note that  observable differences ($\sigma^*=0.03 \sim 0.1$) of positions appear at $t=T^*$.    From then on,   the standard deviation $\sigma_{i,j}(t)$ of position becomes so large ($\sigma_{i,j}> \sigma^*$) that accurate prediction  becomes  impossible, as shown in Fig.~\ref{Fig:transition}.   In other words,  when $t\leq T^*$,  one can give accurate enough prediction about the orbits, but after $t>T^*$, the inherent micro-level physical uncertainty transfers into macroscopic ones ($\sigma_{i,j}> \sigma^*$) so that any ``accurate predictions'' about the orbits of the chaotic three-body system have {\em no} physical meanings at all.  Thus,  $T^*$  gives  the  maximum  time  of theoretical prediction,  called the physical limit of prediction time.   Therefore, when $t < T^*$, although the inherent physical uncertainty propagates exponentially,  accurate enough prediction of orbits is {\em possible} in theory.  However, when $t > T^*$, the micro-level inherent physical uncertainty due to the fluctuation ${\bf r}'_i(0)$ of initial position is enlarged to be macroscopic, say,  the problem becomes essentially random in observation,  as shown in Fig.~\ref{Fig:transition},  so that  it is impossible {\em  in physics} to give any accurate predictions of orbits after the critical time $T^*$.

It should be emphasized that such kind of macroscopic uncertainty comes {\em solely} from the micro-level inherent physical uncertainty due to the fluctuation ${\bf r}'_i(0)$ of initial position, and has nothing to do with human being and Heisenberg's uncertainty principle, since the numerical noises are negligible for each simulations of chaotic orbits (because of the use of the CNS) and besides the model equation is good enough due to  much  smaller   body's  velocities   than  the light speed.  Therefore,  the origin of this kind of macroscopic randomness is the micro-level inherent physical uncertainty of position due to the wave-particle duality of de Broglie and/or the  Planck length  based on the string theory.  Thus, {\em without} any {\em external} disturbance,  the micro-level inherent physical uncertainty  {\em itself}   can be enlarged exponentially and {\em excited} into macroscopic randomness.   Such kind of uncertainty is called {\em self-excited macroscopic uncertainty} or {\em self-excited randomness} due to chaos.   This is a new concept, which can be used to explain the origin of uncertainty/randomness of many phenomena in nature, such as turbulent flows.

\begin{figure}[t]
\begin{center}
\includegraphics[scale=0.3]{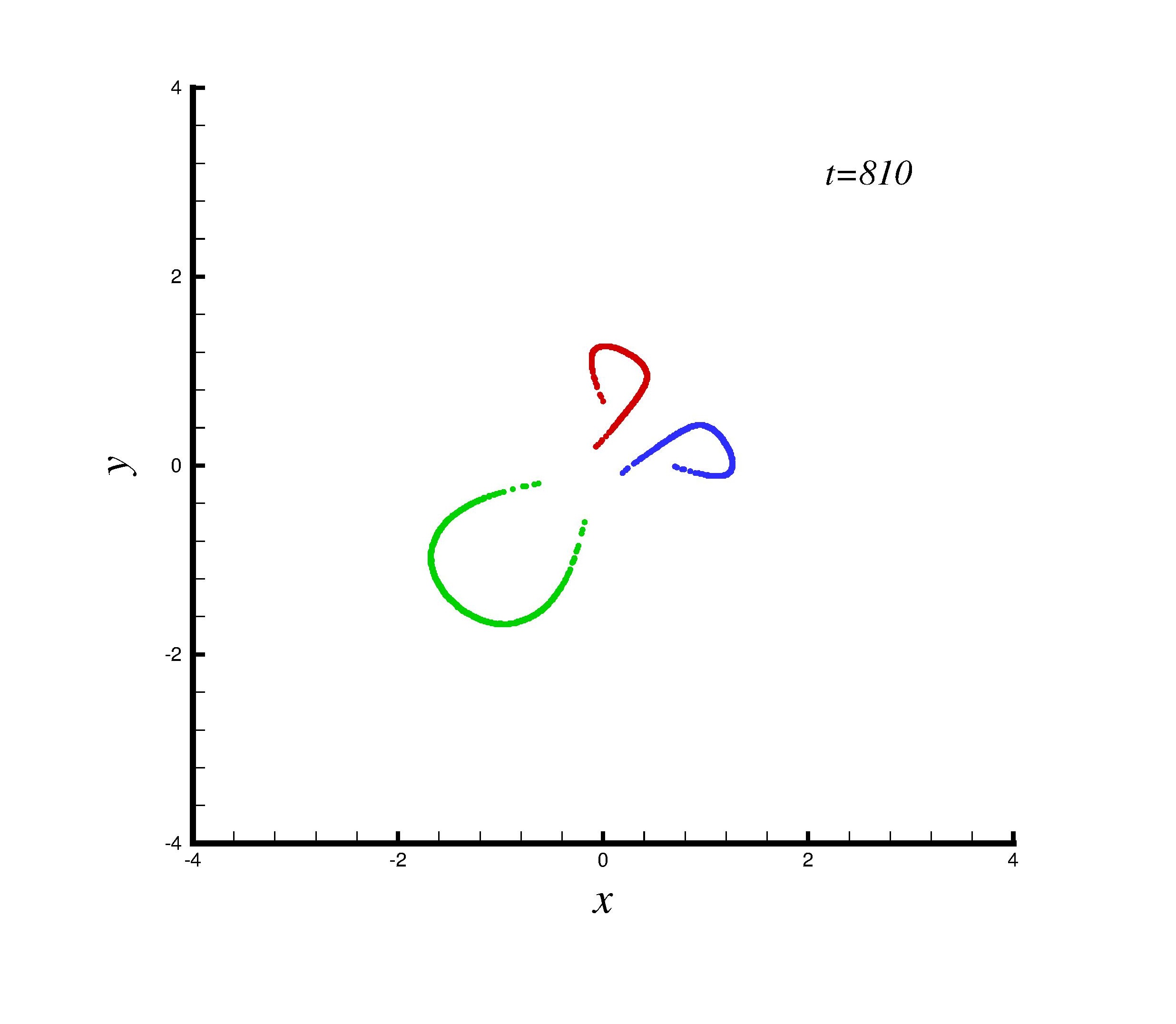}  \includegraphics[scale=0.3]{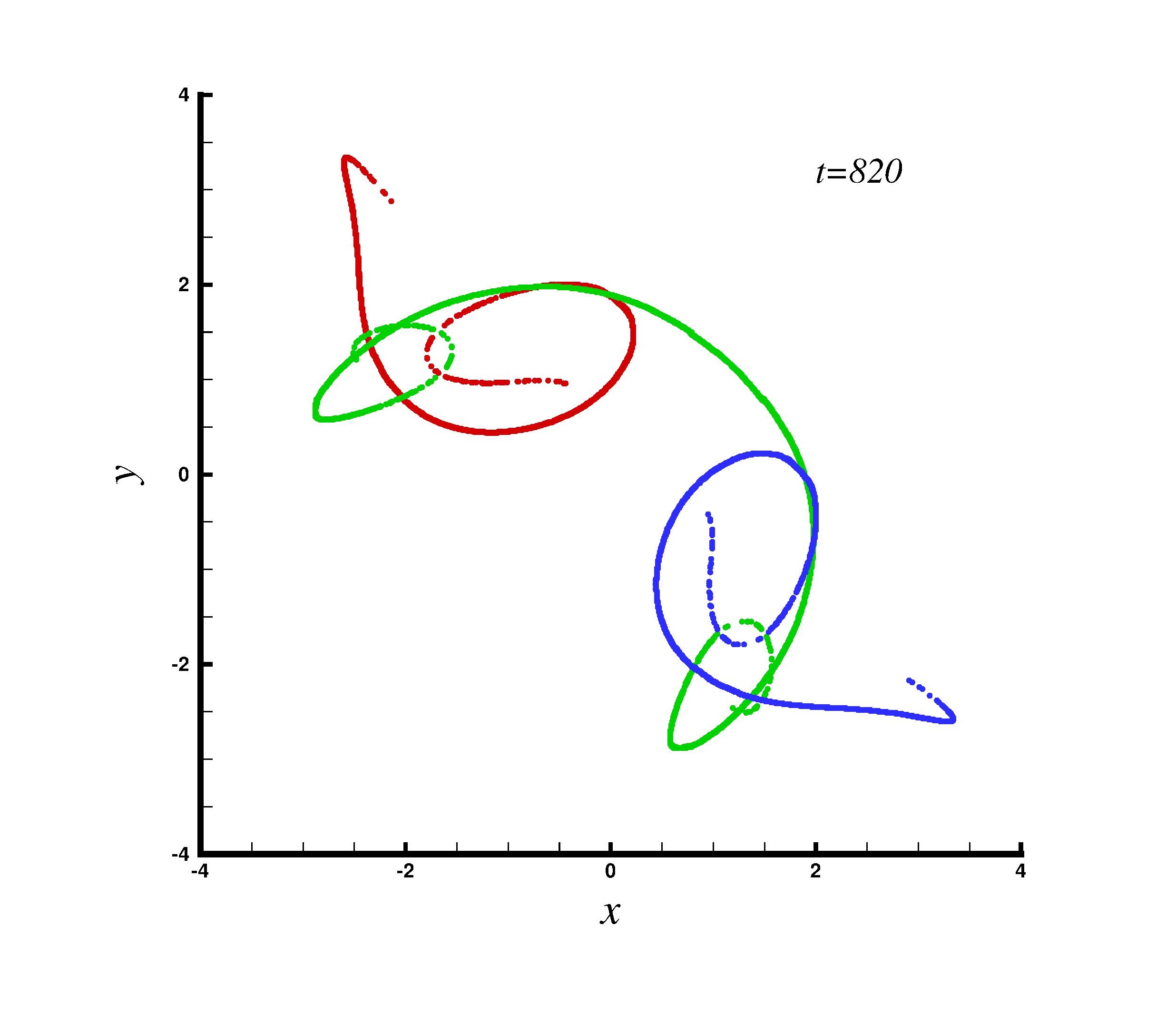} \\
\includegraphics[scale=0.3]{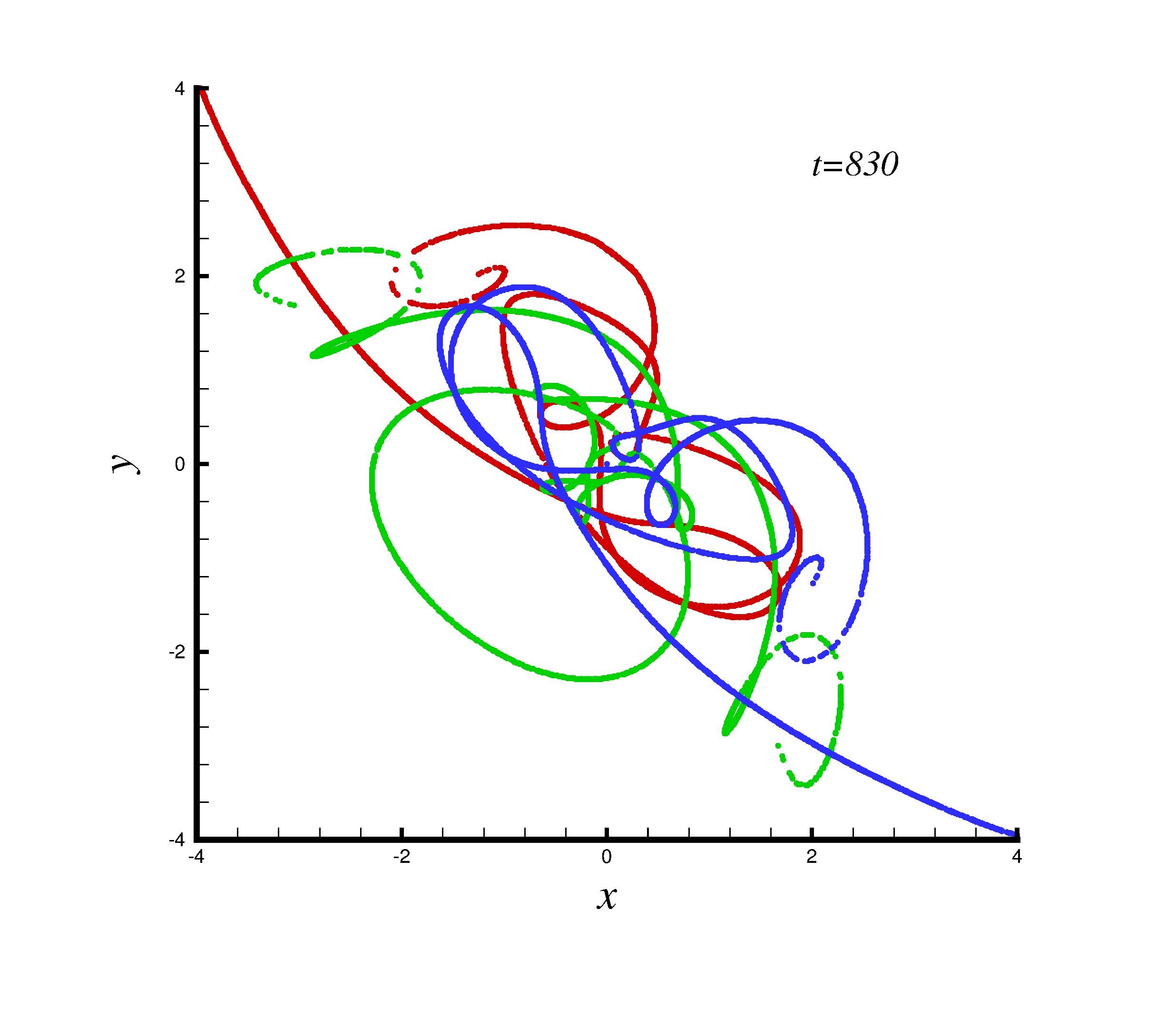}  \includegraphics[scale=0.3]{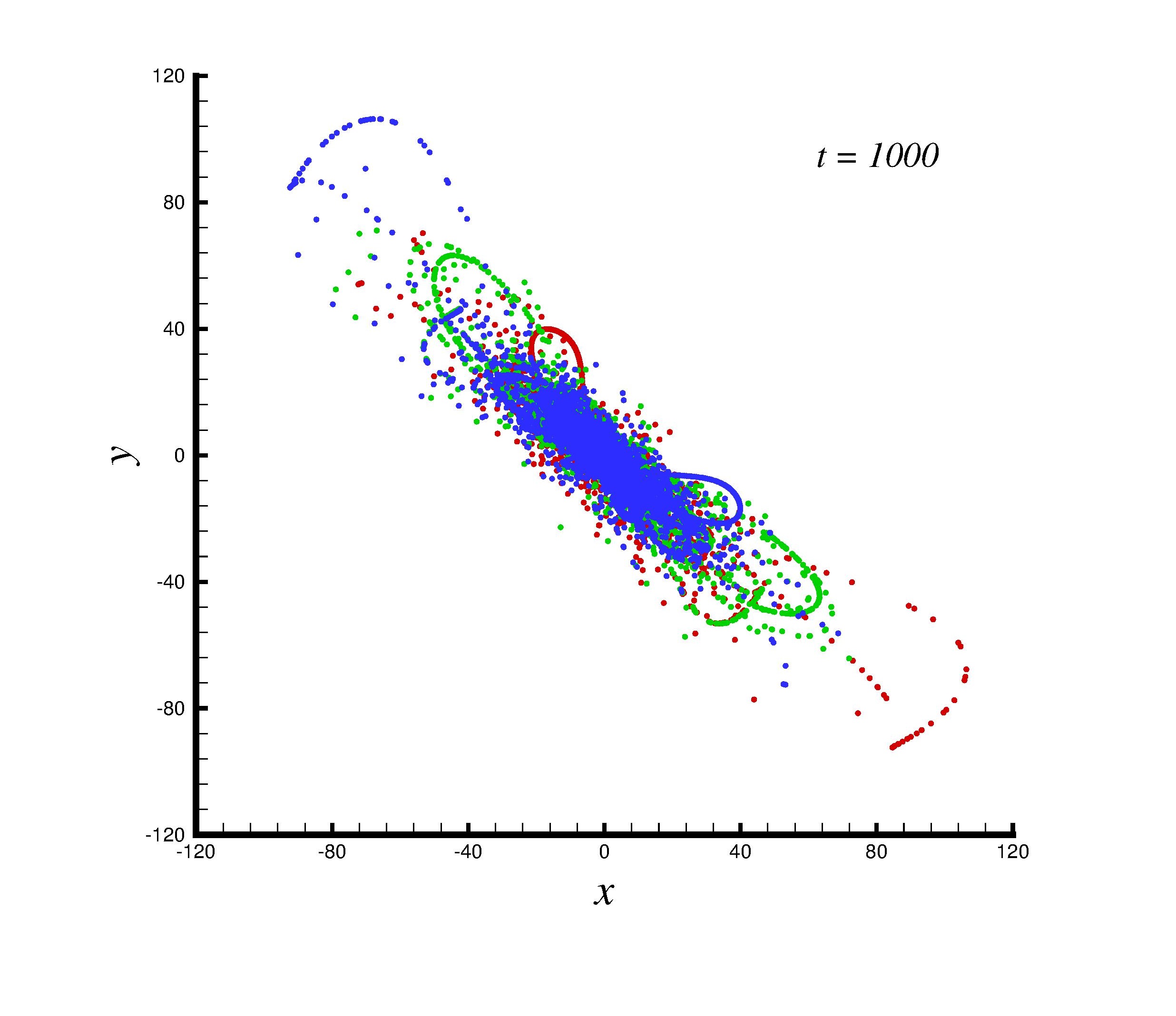}
\caption{{\bf The position distribution of Body-1 (red points), Body-2 (green points) and Body-3 (blue points) in the $(x,y)$ plane at different times when $\sigma_0 =3\times 10^{-60}$.}  Results are  based on the 10000  samples of reliable, multiple-scale simulations given by the CNS with the micro-level fluctuation of initial position ${\bf r}'_i(0)$ in Gaussian distribution.  }
\label{Fig:transition:B}
\end{center}
\end{figure}

It is found that the three-body system does not disrupt even in the interval $[0,10000]$ if there is {\em no} micro-level inherent physical uncertainty of positions in the initial conditions, i.e. ${\bf r}'_i(0)=0$,  mainly due to the symmetry of the initial condition.   However, it is very interesting that the tiny, micro-level, physical  fluctuation of position with the initial standard deviation $\sigma_0=10^{-60}$ might  lead to a totally different destiny of the three-body system:  when the inherent physical uncertainty is enlarged into macroscopic,  2568  among 10000  samples  of the three-body system  disrupt  at $t=1000$, i.e.  with  one  body  escaping  in a random direction and the other two becoming binary stars in the opposite direction.  Thus, the micro-level physical uncertainty due to initial position fluctuation  ${\bf r}'_i(0)$, although it is rather tiny, can greatly influence the destiny of the chaotic three-body system.  It should be emphasized that these 10000 {\em mathematically} different micro-level fluctuations ${\bf r}'_i(0)$ of position are the {\em same} for us from the {\em physical} viewpoint, since a  spatial difference  {\em less} than the Planck length has no {\em physical} meaning at all.  However,   these {\em physically  same} initial conditions lead to   completely {\em different}  orbits  and  even  {\em different}  destiny  of  the  chaotic  three-body  system!   Note that, whether the three-body system disrupts at $t=1000$ or not depends upon the micro-level {\em inherent} physical fluctuation ${\bf r}'_i(0)$ of position in Gaussian distribution with the standard deviation $10^{-60}$.   It should be emphasized that such kind of disruption of the three-body system {\em randomly} happens  {\em without}  any {\em external} disturbances.   This phenomena  is called the {\em self-excited random disruption} or {\em self-excited random escape} of three-body system.   It suggests that a chaotic three-body system would {\em randomly} evolve {\em by itself}  to a rather complicated structure {\em without} any {\em external} forces.   It also implies that  an universe could  {\em randomly} evolve by {\em itself} into complicated structures,  {\em without} any external forces, and that the nature could {\em randomly} evolve by itself into organism and even human being,  {\em without} any external forces.

\begin{figure}[t]
\begin{center}
\includegraphics[scale=0.30]{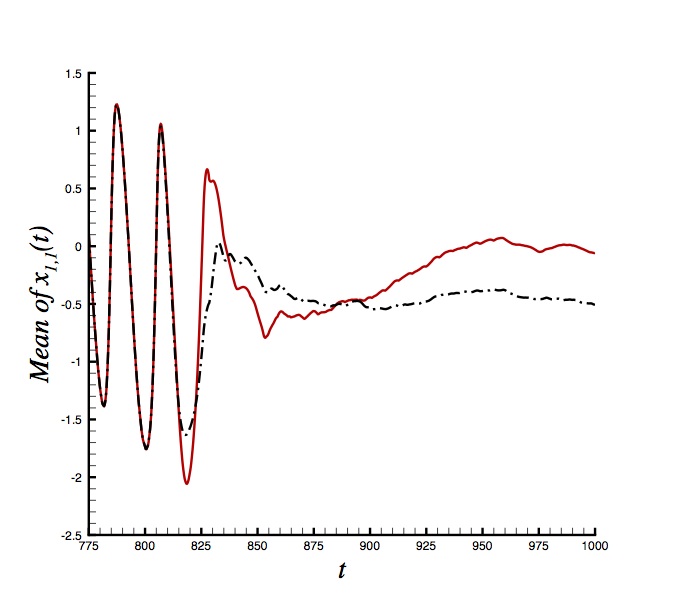}
\includegraphics[scale=0.30]{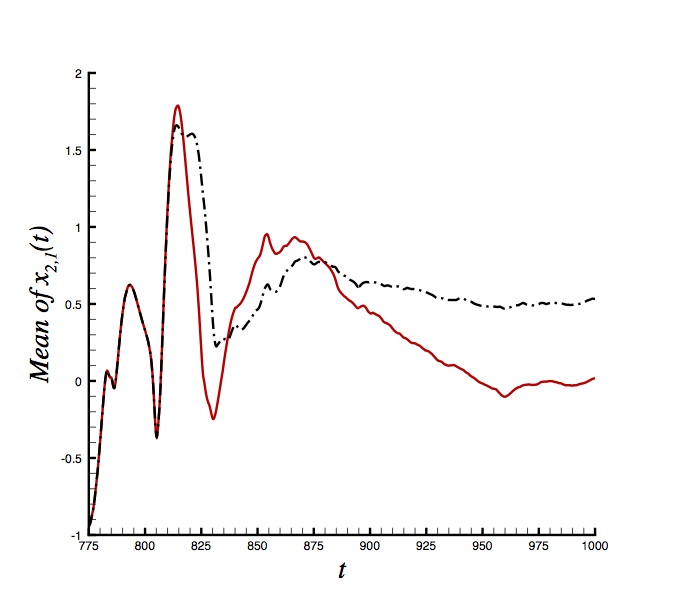}
\includegraphics[scale=0.30]{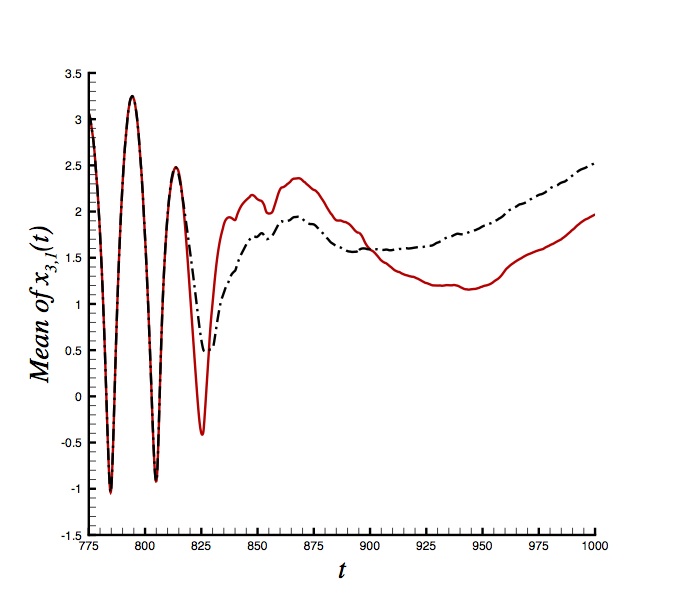}
\caption{{\bf Comparison of the mean position $(\overline{x}_{1,1}, \overline{x}_{2,1}, \overline{x}_{3,1})$ of Body-1 given by $\sigma_0=10^{-60}$ and $\sigma_0=3\times 10^{-60}$.}  Each curve is  based on the 10000  samples of reliable, multiple-scale simulations given by the CNS with the micro-level fluctuation of initial position ${\bf r}'_i(0)$ in Gaussian distribution (statistic results given by the 10,000 samples are the same as those given by 8000 ones).  Solid line:  statistical result in case of $\sigma_0=10^{-60}$;  Dashed line: statistical result in case of  $\sigma_0=3\times 10^{-60}$.  }
\label{Fig:compare-sigma}
\end{center}
\end{figure}

All of the above-mentioned results are based on 10000 samples of  reliable (convergent)  numerical simulations of  the chaotic three-body system (under consideration)  by means of the CNS using the inherent micro-level physical fluctuation of initial position with $\sigma_0 = 10^{-60}$.  Similarly, it is straightforward to investigate the propagation of  the uncertainty of positions  of the chaotic three-body systems for other values of $\sigma_0$.  Without loss of generality,  let us consider the case with the same mean position (\ref{mean-position:0}) and velocity (\ref{mean-velocity:0}), but different fluctuation of initial position in Gaussian distribution:
 \[   \left<{\bf r}'_i(0)\right> = 0, \;\;\;      \sigma_0 = \sqrt{\left<{\bf r}'^2_i(0)\right>} =3.0 \times 10^{-60}. \]
 It is found that the corresponding $\sigma_{i,j}(t)$ of $x_{i,j}(t)$ also enlarges exponentially until $T^*\approx 810$, the  so-called  physical  limit    of  prediction time, and then propagates {\em algebraically} thereafter.  When $t<T^*$,  $\sigma_{i,j}(t)$ is so small that accurate prediction of orbits is possible, although it enlarges exponentially in the same way $\sigma_{i,j}(t) = \sigma_0 \exp(\lambda t)$, where $\lambda=0.1681$ is the Lyapunov exponent  given by \cite{Sprott2010} for the same three-body system without fluctuation of initial position.   However, when $t>T^*$, the uncertainty becomes macroscopic and observable,  as shown in Fig.~\ref{Fig:transition:B}.
 Comparing Fig.~\ref{Fig:transition} with Fig.~\ref{Fig:transition:B}, it is obvious that the macroscopic statistical distributions of position of the three-body system at different times are dependent upon $\sigma_0$, i.e. the standard deviation of the micro-level physical inherent fluctuation of initial position ${\bf r}_i(0)$.    Note that,  when $t>T^*$,  the mean positions of the chaotic three-body system  given by different $\sigma_0$  depart obviously, as shown in Fig.~\ref{Fig:compare-sigma}.    This suggests that the statistics of the macroscopic uncertainty of the chaotic three-body system have a close relationship with the statistics of  the micro-level inherent physical uncertainty.     In addition,    2736 ``random disruptions'' and ``random escapes''  (among 10000 samples)  happen  in the time interval from $t=T^*  \approx 810$  to  $t=1000$,  {\em without} any {\em external} disturbance,  which is about 1.7\%  higher than that in case of $\sigma_0 =3 \times 10^{-60}$.    So,  it seems that the  percentage of this kind of random escape and disruption at $t > T^*$ is dependent upon the micro-level  physical inherent uncertainty of  the initial positions and velocities of the three-body system.   In other words,  the {\em macroscopic} statistic might be {\em dependent} upon  the {\em microscopic} physical uncertainty of the initial condition of the chaotic three-body system under consideration!           
 
 Note that, when the micro-level inherent  physical uncertainty is not considered, i.e. ${\bf r}'_i(0)=0$, the Body-2 moves along a straight line, and Body-1,  Body-3 have the symmetry to the Body-2.  However, when ${\bf r}'_i(0)\neq 0$, such kind of symmetry breaks  {\em randomly}   after  $t= T^* \approx 810$ when the micro-level physical uncertainty transfers into the macroscopic randomness,  {\em without any external disturbance}!   Such kind of  ``random''  symmetry breaking of the three-body system is called ``self-excited symmetry breaking''.      
 
 In the theory of chaos,  Lorenz's ``butterfly-effect'' is very famous, say,  a hurricane in North America might be created by the flapping of the wings of a distant butterfly  in South America several weeks earlier.  Note that, the flapping of the wings of a butterfly is a kind of {\em external}  disturbance.  Besides, such kind of external disturbance is much larger and stronger than the micro-level  inherent  physical fluctuations.  However, in our simulations mentioned above,  no  ``external'' disturbance exists at all, since micro-level physical fluctuation of the initial position is {\em inherent} and {\em inside} the system!   So, different from the famous ``butterfly effect''of chaos, our computations reveal a kind of ``molecule-effect'' of chaos, say, a hurricane in North America might be created even by  a random motion of a  distant  molecule  of  the  air in South America  several weeks earlier!   Thus, a hurricane could be created even {\em without} flapping of the wing of any butterflies.    This ``molecule-effect'' of chaos reveals more deeply  the essence of the so-called ``sensitive dependence  on initial condition'' (SDIC) of chaotic dynamic systems.   In other words, {\em without any external forces}, the considered 3-body  chaotic system  can evolve itself into escape, disruption and symmetry breaking!   And a hurricane in North America might be created no matter whether a distant butterfly  in South America  flaps or not!  Therefore,  this  kind  of  ``molecule-effect'' of chaos can be regarded as  a  ``non-butterfly effect'' of chaos, because a chaotic dynamic system (like the 3-bodies considered in this paper) is  inherently random in physics.   From this view of point,  it has no meaning to say that ``a chaotic dynamic system is deterministic'', or ``a deterministic chaotic system might lead to random''.    This can explain the origins of many complicated phenomena such as turbulent flows.

 \section{Concluding remarks}

In summary, the  microscopic inherent uncertainty (in the level of $10^{-60}$)  due to physical  fluctuation of initial positions of the three-body system  enlarges {\em exponentially}  into  macroscopic  randomness (at the level $O(1)) $ until $t=T^*$,  the so-called physical limit  of prediction time,  but propagates {\em algebraically}  thereafter  when accurate prediction of orbit is impossible.   Note that these 10000 samples use micro-level, inherent physical fluctuations of initial position, which have nothing to do with human being.   Especially, the differences  of  these 10000  fluctuations are {\em mathematically} so small (in the level of $10^{-60}$ ) that all of these initial conditions are {\em physically}  the same since a spatial difference  shorter than a Planck length does  not make physical senses according to the spring theory \cite{Davies1988}.    It indicates that the macroscopic randomness  of the chaotic three-body system is {\em self-excited}, say, {\em without} any {\em external} force or disturbances, from the  micro-level inherent physical uncertainty.  This provides us the new concept ``self-excited macroscopic randomness/uncertainty''.   It is found that  the macroscopic randomness is even dependent upon microscopic uncertainty, from statistical viewpoint.   Besides, it is found that  the chaotic three-body system might {\em randomly}  disrupt  at $t=1000$ in about 25\%  probability {\em without} any external disturbance,  which provides us the new concepts ``self-excited random disruption'' and ``self-excited random escape'' of chaotic three-body system.   In addition, the symmetry of motion of this chaotic 3-body system begins to {\em randomly} break at $t=T^* \approx 810$, {\em without} any external disturbance, which provides us the new concept ``self-excited symmetry breaking''.   All of these  suggest  that a chaotic three-body system might {\em randomly} evolve by itself into escape, disruption and symmetry-breaking, {\em without} any external forces or disturbance.     Thus, the world is essentially uncertain,  since such kind of self-excited  macroscopic randomness/uncertainty,  self-excited escape/disruption and self-excited symmetry-breaking are inherent and unavailable.   This work also implies that  an universe could  {\em randomly} evolve by {\em itself} into complicated structures {\em without} any {\em external} forces, and similarly that the nature could {\em randomly} evolve by {\em itself} into organism and even human being,  {\em without} any external forces!

In this paper,  the SDIC of chaos is considered from  a  new  viewpoint of physics.   Especially,  the  micro-level  physical  fluctuations  of  initial positions of the three-body system are so small (at the level of $10^{-60}$) that the initial conditions of these 10,000 samples are {\em mathematically} different but   {\em physically}  the same, since a spatial difference shorter than a Planck length does  not make physical senses according to the spring theory \cite{Davies1988}.  So, {\em physically} speaking, there is {\em no} (external) disturbance at all at the initial condition!  However, it is very interesting that, the 10,000 samples with the {\em physically}  {\bf same} initial conditions evolve into  completely  {\bf different}  trajectories, even when  the numerical noise is negligible.  This is quite different from the traditional ``butterfly-effect'' that emphasizes  the sensitive dependance of the chaotic trajectories on {\em physically different} initial conditions  caused  by  {\em external}  disturbance.  To emphasize this kind of difference,  the so-called  ``molecule-effect'' or ``non-butterfly effect'' of chaos is suggested in this paper, which emphasizes the sensitive dependence of the chaotic trajectories on {\em physically same} initial conditions  {\em without} any  {\em external}  disturbance.   To the best of our knowledge,  these results have never been reported.   This is mainly because quantitative variation can lead to qualitative change:  the very high accuracy of the CNS \cite{Liao2009, Liao2013, Liao2013-3b, LiaoWang2014, XiaoMingLi-2014}  greatly deepens our understandings about the SDIC of chaos and enriches our knowledge about the essence of chaos, from a  new viewpoint of physics.     

Note that, due to the SDIC of chaos, chaotic results given by the traditional numerical methods (such as Runge-Kutta method) are mixtures of true trajectories and numerical noises, so that it is {\em impossible} to investigate the propagation of the micro-level physical uncertainty of initial conditions which are much smaller than numerical noises.  However, by means of the CNS  \cite{Liao2009, Liao2013, Liao2013-3b, LiaoWang2014, XiaoMingLi-2014}, the numerical noises can be reduced greatly so that they are much smaller than the physical micro-level uncertainty.  By means of the CNS,  we are now quite sure that the micro-level physical uncertainty of the initial conditions can transfer into macroscopic randomness and besides even the macroscopic statistic results are dependent upon the micro-level physical uncertainty.  As mentioned above, the  reliable (convergent) simulations of the chaotic 3-body system given by the CNS  {\em more profoundly} reveal the essence of the SDIC of chaos  from  a  totally new viewpoint of physics.  Besides, using the CNS and National Supercomputer TH-1A, the reliable chaotic simulation of Lorenz equation in a very long interval [0,10000] of time was obtained \cite{LiaoWang2014} that is helpful to stop the intense arguments about chaos \cite{Liao2014-IJBC}.  In addition, it is found \cite{XiaoMingLi-2014} by means of the CNS that at least seven among the currently reported 15 periodic orbits of a three-body system \cite{3body-2012} greatly depart from the periodic ones within a long enough interval of time, and are thus most possibly unstable.   Therefore, the CNS  \cite{Liao2009, Liao2013, Liao2013-3b, LiaoWang2014, XiaoMingLi-2014} indeed provides us a useful tool to investigate the chaos more accurately.      

More importantly, all of these reliable computations reveal an origin of macroscopic randomness and uncertainty, i.e. the micro-level uncertainty, which might be exponentially enlarged into macroscopic randomness and uncertainty due to chaos.   In other words, {\em chaos might be a bridge between the micro-level uncertainty and macroscopic randomness}.     

\section*{Acknowledgement}
The calculations of this work were performed on TH-1A  at National Supercomputer Center in Tianjin, China.  It is partly supported by National Natural Science Foundation of China (Approval No.  11272209 and 11432009)  and State Key Lab of Ocean Engineering (Approval No. GKZD010063).

\bibliography{3body}
\bibliographystyle{unsrt}

\end{document}